\documentstyle[11pt,newpasp,twoside, epsf]{article}
\def\edcomment#1{\iffalse\marginpar{\raggedright\sl#1\/}\else\relax\fi}
\reversemarginpar

\def\ltsima{$\; \buildrel < \over \sim \;$}
\def\simlt{\lower.5ex\hbox{\ltsima}}
\def\gtsima{$\; \buildrel > \over \sim \;$}
\def\simgt{\lower.5ex\hbox{\gtsima}}

\begin{document}
\title{Maps of the millimetre sky \\ from the BOOMERanG experiment}

\author{
P.~de~Bernardis$^{1}$, P.A.R.~Ade$^{2}$, A. Balbi$^{3}$,
J.J.~Bock$^{4}$, J.R.~Bond$^{5}$, J.~Borrill$^{6}$,
A.~Boscaleri$^{7}$, P. Cabella$^{3}$, C.R.~Contaldi$^{5}$,
B.P.~Crill$^{8,9}$, G. De Gasperis$^{3}$, A. De-Oliveira
Costa$^{10}$, G.~De~Troia$^{1}$, K.~Ganga$^{11}$,
M.~Giacometti$^{1}$, E.~Hivon$^{11}$, V.V.~Hristov$^{8}$, T.
Kisner$^{12}$, A.~Iacoangeli$^{1}$, A.H.~Jaffe$^{13}$,
W.C.~Jones$^{8}$, A.E.~Lange$^{8}$, S.~Masi$^{1}$, P.~Mason$^{8}$,
P.D.~Mauskopf$^{2}$, C. MacTavish$^{14}$, A.~Melchiorri$^{1}$,
T.~Montroy$^{11}$, F. Nati$^{1}$, P. Natoli$^{3}$,
C.B.~Netterfield$^{14}$, E.~Pascale$^{7,14}$, F.~Piacentini$^{1}$,
D.~Pogosyan$^{15}$, G.~Polenta$^{1}$, S.~Prunet$^{16}$,
G.~Romeo$^{17}$, S. Ricciardi$^{1}$, J.E.~Ruhl$^{12}$, M.
Tegmark$^{10}$, N. Vittorio$^{3}$}

\affil{
$^{1}$ Dipartimento di Fisica, Universita' La Sapienza,
        Roma, Italy \\
$^{2}$ Dept. of Physics and Astronomy, Cardiff University, Wales, UK \\
$^{3}$ Dipartimento di Fisica, Universita' di Tor Vergata, Roma, Italy \\
$^{4}$ Jet Propulsion Laboratory, Pasadena, CA, USA \\
$^{5}$ CITA, University of Toronto, Canada \\
$^{6}$ NERSC, LBNL, Berkeley, CA, USA \\
$^{7}$ IFAC-CNR, Firenze, Italy \\
$^{8}$ California Institute of Technology, Pasadena, CA, USA \\
$^{9}$ CSU Dominguez Hills, Carson, CA, USA \\
$^{10}$ Phys. Dept. University of Pennsylvanya, Philadelphia, PA, USA \\
$^{11}$ IPAC, CalTech, Pasadena, CA, USA \\
$^{12}$ Physics Department, CWRU, Cleveland, OH, USA \\
$^{13}$ Astrophysics Group, Imperial College, London, UK \\
$^{14}$ Depts. of Physics and Astronomy, University of Toronto, Canada \\
$^{15}$ Physics Dept., University of Alberta, Alberta, Canada \\
$^{16}$ Institut d'Astrophysique, Paris, France \\
$^{17}$ Istituto Nazionale di Geofisica, Roma,~Italy \\ }

\begin{abstract}
In the 1998-99 flight, BOOMERanG has produced maps of $\sim 4 \%$
of the sky at high Galactic latitudes, at frequencies of 90, 150,
240 and 410 GHz, with resolution $\simgt 10'$. The faint structure
of the Cosmic Microwave Background at horizon and sub-horizon
scales is evident in these maps. These maps compare well to the
maps recently obtained at lower frequencies by the WMAP
experiment. Here we compare the amplitude and morphology of the
structures observed in the two sets of maps. We also outline the
polarization sensitive version of BOOMERanG, which was flown early
this year to measure the linear polarization of the microwave sky
at 150, 240 and 350 GHz.
\end{abstract}

\section{Introduction}

BOOMERanG is balloon-borne microwave telescope, sensitive at 90,
150, 240 and 410 GHz, with a resolution of $\sim 10'$. The
instrument was equipped with very sensitive spider-web bolometers
(Mauskopf et al. 1997). The image of the sky is obtained by slowly
scanning the full payload in azimuth ($\pm 30^o$) at constant
elevation. The scan center constantly tracks the azimuth of the
lowest Galactic foreground region, situated in the southern
hemisphere, in the Horologium constellation. Every day of the
flight the instrument obtains a fully cross-linked map of about
$45^o \times 30^o$ of the sky. The instrument was flown in a long
duration circum-Antarctic flight from Dec.28, 1998 to Jan.8, 1999,
and has been described in Piacentini et al. (2002) and in Crill et
al. (2003). The main target of the experiment was the detection of
anisotropy in the Cosmic Microwave Background. The maps produced
by the experiment have been published in de Bernardis et al.
(2000), Masi et al. (2001), Netterfield et al. (2002), Ruhl et al.
(2003). Due to the limited size and to 1/f noise, these maps do
not contain information for angular scales $\gg 5^o$ and have been
filtered accordingly. The maps are calibrated to $\sim 10 \%$ in
gain and to $\sim 10\%$ in beam FWHM. In the 150 GHz map, the
signal is well above the noise, and maps taken at different scan
speeds and in different locations are perfectly consistent,
demonstrating the low level of systematic effects. The level of
the noise is of the order of 50 $\mu K$ per $7'$ pixel and of 20
$\mu K$ per $28'$ pixel. At 90, 150 and 240 GHz the rms signal has
the spectrum of CMB anisotropy, and does not fit any reasonable
spectrum of foreground emission. The temperature fluctuations
detected in the high latitude part of the 150 GHz map are
remarkably gaussian (Polenta et al. 2002, De Troia et al. 2003).
The angular power spectrum of the 150 GHz map in the multipole
range $50 < \ell < 1000$ has been computed in Netterfield et al.
(2002) and Ruhl et al. (2003). Three peaks have been detected in
the power spectrum, at multipoles $\ell \sim$ 210, 540, 845 (de
Bernardis et al., 2002). Interstellar Dust contamination of the
150 GHz power spectrum has been shown to be less than 1\% (Masi et
al. 2001). These results fit the scenario of acoustic oscillations
of the primeval plasma at horizon and subhorizon scales (Sunyaev
and Zeldovich 1970, Peebles and Yu 1970). In the framework of
adiabatic inflationary structure formation the cosmological
parameters have been estimated from the measured power spectrum
(Lange et al. 2001, Netterfield et al., 2002, de Bernardis et al.
2002, Ruhl et al. 2003). The three cosmological parameters best
constrained by the BOOMERanG data are the curvature parameter
$\Omega = 1.03\pm 0.05$  (the universe is nearly flat), the
spectral index of the primordial density perturbations $n_s =1.02
\pm 0.07$ (nearly scale invariant), the physical density of
baryons $\Omega_b h^2 = 0.023 \pm 0.003$ (consistent with Big Bang
Nucleosynthesis).

Here we compare the BOOMERanG maps to the maps recently obtained
at similar frequency and resolution by the WMAP satellite (Bennett
et al., 2003). Working from the advantage L2 point of the
Sun-Earth system, in its first year of operation WMAP has produced
full sky maps of the microwave sky at 22, 32, 41, 60, 94 GHz, with
resolution of the order of $15'-30'$ and noise of $\sim 35 \mu K$
per pixel in 28' pixels. The maps are precisely calibrated (better
than 1\%). The power spectra obtained from these maps are fully
consistent with the CMB power spectrum measured by BOOMERanG at
150 GHz, and with the adiabatic inflationary scenario: they allow
the precise determination of most of the cosmological parameters;
these estimates improve the precision of the BOOMERanG ones by a
factor 2-3 for the parameters mentioned above; the full-sky
coverage and polarization sensitivity allows the determination of
previously poorly constrained parameters, like the reionization
optical depth (Spergel et al. 2003).

\noindent The comparison of the BOOMERanG and WMAP maps is carried
out with three targets:

\noindent $\bullet$ Compare independent maps of the CMB with
similar angular resolution, and confirm the detection of
primordial structures in both experiments;

\noindent $\bullet$ Improve the BOOMERanG calibration using the
precise calibration of WMAP, and demonstrate that 1\% gain
calibration is possible for the new, polarization-sensitive
BOOMERanG-B2K survey;

\noindent $\bullet$ Infer the level and properties of foregorund
contamination in forthcoming deep surveys of the CMB.

\section{BOOMERanG vs WMAP}

In fig. 1 we compare the maps of BOOMERanG at 90, 150, 240 GHz to
the maps of WMAP at 41, 60, 94 GHz in the same high latitude
region. The maps are Healpix (Gorski et al. 1998) representations
with npix=512 (7' per pixel). The region selected has the best
coverage in the BOOMERanG 150 GHz channels. A $ \sim 5^o$
high-pass has been applied to the WMAP maps to easy the comparison
to the BOOMERanG maps, which are intrinsically insensitive to
large angular scales.

The morphological agreement of the structures detected is evident.
Significant differences are in the level of the noise and in the
presence of important contamination by AGNs in the lower frequency
channels. Once these sources are masked, the pixel-pixel
correlation between the maps is quite good.

\begin{figure}
\plotone{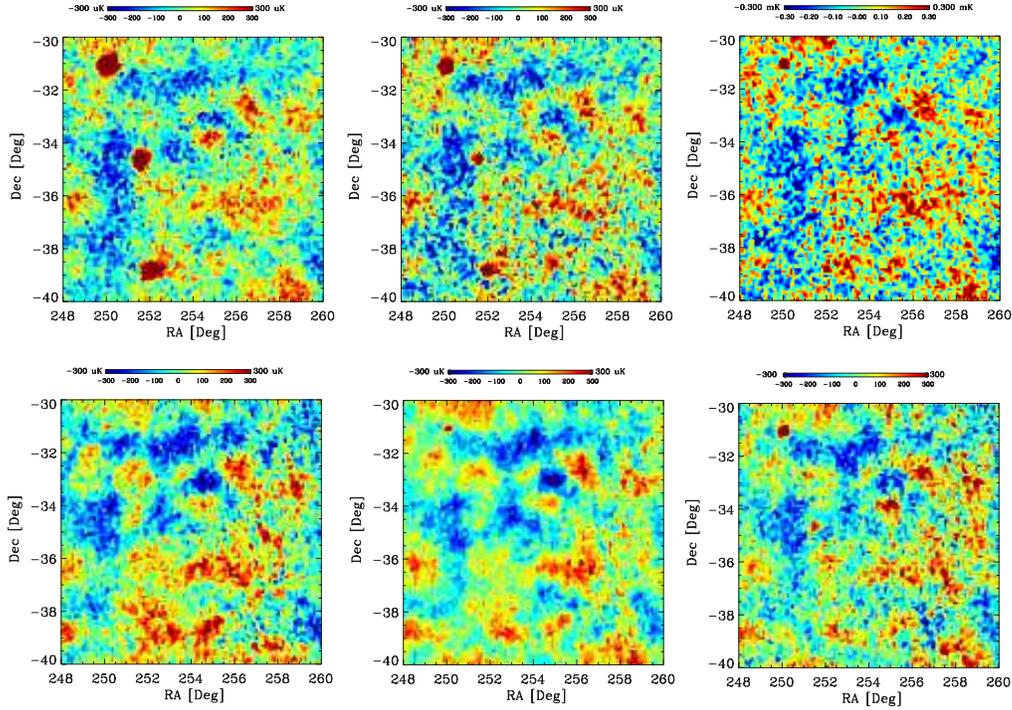} \caption{A sample $\sim 10^o \times 10^o$
region at high galactic latitudes as seen by WMAP (top row, 41,
61, 94 GHz left to right) and by BOOMERanG-98 (bottom row, 220,
150, 90 GHz left to right). The grey scale is in thermodynamic
temperature units for the CMB ($-300 \mu K < \Delta T < 300 \mu
K$). The coordinates are Galactic ($248^o < \ell <260^o$, $-40^o <
b < -30^o$). The pixel size is 7'. It is evident the very good
agreement of the CMB maps obtained by the two experiments (compare
the two $\sim$ 90 GHz maps on the right). Note the decrease of the
equivalent brightness of the three AGN sources (evident at 41 GHz)
with increasing frequency, and the very low noise of the 150 GHz
map by BOOMERanG (center panel in bottom row). The WMAP maps have
been filtered to remove structures larger than 5$^o$, which cannot
be detected by BOOMERanG.}
\end{figure}

A quantitative analysis must take into account all important
differences between the two datasets:

\noindent $\bullet$ The beams of BOOMERanG and WMAP are different.
The BOOMERanG beam at 150 GHz is closely fit by a 11' gaussian
down to 2\% of the axial gain, while at lower levels is better fit
by a 13' gaussian. The WMAP beam at 94 GHz is fit by a 13'
gaussian down to 5\% of the axial gain, and has wide "shoulders"
at lower levels. This difference is difficult to treat in pixel
space, while its treatment is relatively simple in multipoles
space. An example of this can be found in Abroe et al. (2003),
comparing MAXIMA and WMAP maps of the CMB.

\noindent $\bullet$ The BOOMERanG maps do not include structures
larger than $10^o$, while WMAP maps are accurate at all scales.
Once again, this is easier to account for in multipoles space.

\noindent $\bullet$ In this pixelization, the noise of the 150 GHz
BOOMERanG map is around 50 $\mu K$ per pixel, while the noise of
the 94 GHz WMAP map is around 180 $\mu K$ per pixel.

A full analysis taking into account all these details is described
in Hivon et al. (2003): the main result is that the calibration of
BOOMERanG-98 at 150 GHz is found to be consistent with the precise
calibration of WMAP within 5\%. Moreover, $< 1\%$ calibration of
BOOMERanG is found to be reachable with this correlation method.

Given the close consistency of the maps of BOOMERanG and WMAP, the
consistency of the angular power spectra measured by the two
experiments is almost a trivial consequence. In fig.2 we compare
the angular power spectra detected by the two instruments. The
gain calibration error ($< 1 \%$ for WMAP and $\sim 10 \%$ for
BOOMERanG) is not included in the error bars, which account for
random errors only. It is evident that the WMAP experiment has
much higher sensitivity than BOOMERanG at multipoles $\simlt 450$:
the WMAP measurement in this range is limited by cosmic variance,
and the Power Spectrum can be considered definitive. In the region
of the second peak the two experiments have similar sensitivity,
while in the region of the third peak the BOOMERanG experiment has
better sensitivity due to the smaller beam and the lower noise per
pixel. For the determination of the cosmological parameters
$\Omega$, $\Omega_b h^2$, $n_s$, in the adiabatic inflationary
scenario, WMAP takes advantage of the accurate calibration, while
the BOOMERanG is still competitive because of the wider multipoles
coverage, including the third peak of the spectrum.
\begin{figure}
\plotone{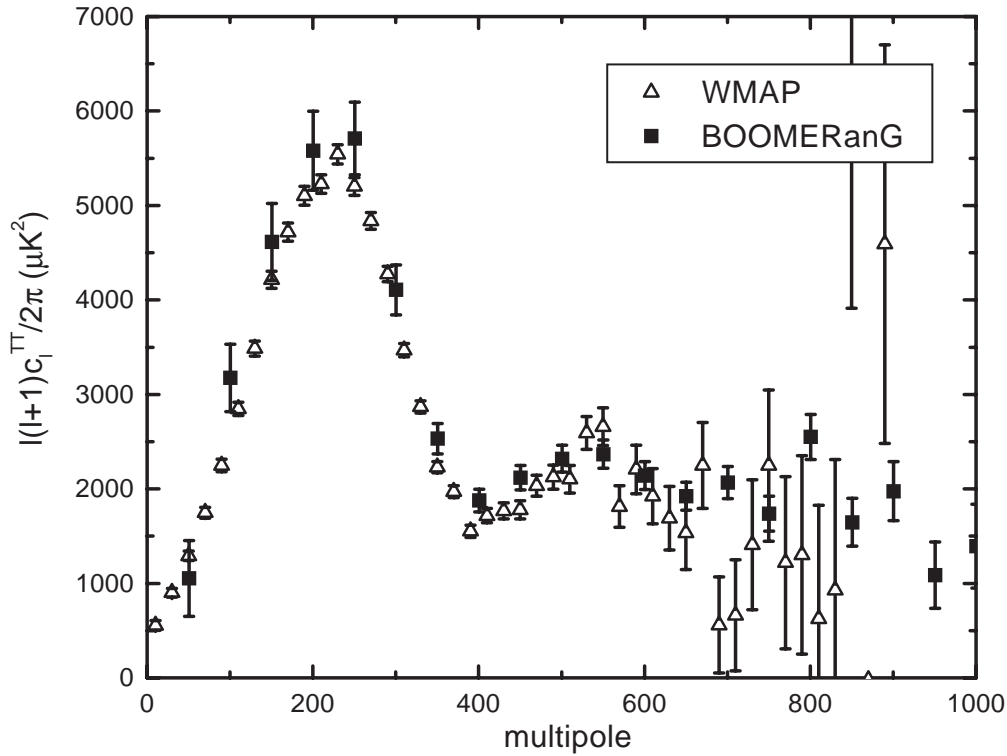} \caption{Comparison of the angular power
spectra of the CMB measured by WMAP (open triangles, $\Delta \ell
= 20$ bins) and by BOOMERanG (squares, $\Delta \ell = 50$ bins).
The error bars represent random errors only.}
\end{figure}
These results give only an idea of what can be expected from
Planck-HFI (Lamarre et al. 2003), the satellite instrument
developed to fully exploit the capabilities of cryogenic
bolometers. This will work for about two years from the same deep
space location as WMAP, using bolometric detectors at 0.1K, even
more sensitive than the BOOMERanG ones.

\section{The polarization-sensitive BOOMERanG: B2K}

After the 1998-99 flight, BOOMERanG has been recovered and
upgraded. Additional attitude sensors have been implemented (a
day-time star camera developed in Toronto and a pointed star
sensor developed in Rome), and the focal plane has been rebuild to
accomodate polarization sensitive bolometers (PSB) developed in
JPL/Caltech (Jones et al. 2003). The new focal plane is sketched
in fig.3.
\begin{figure}
\plotone{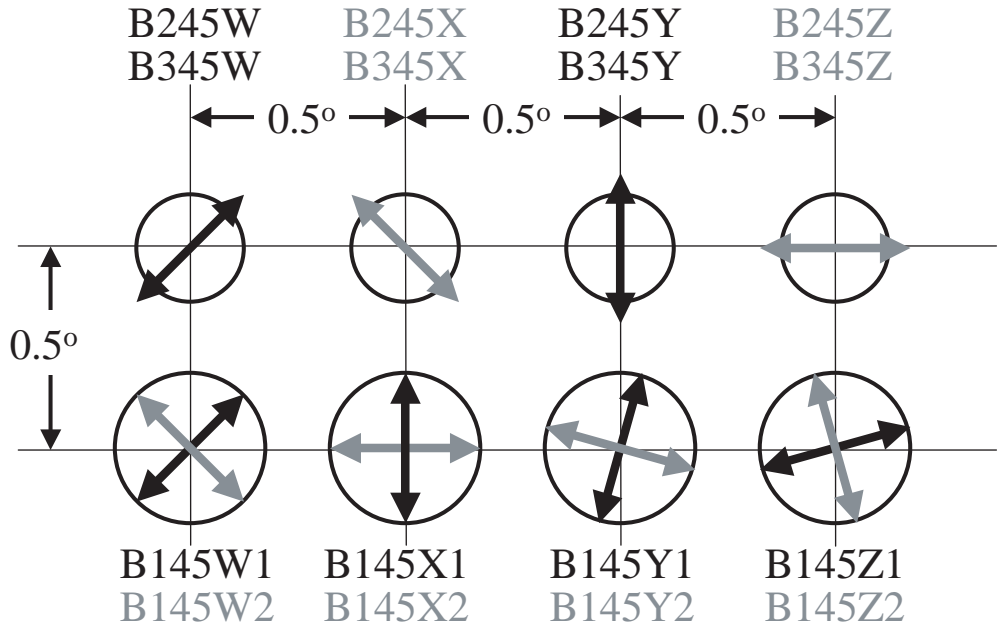} \caption{The focal plane of B2K. The azimuth
(scan) direction is horizontal. The 8 pixels host two bolometers
each, labeled by the frequency in GHz and by the principal axis of
the polarization sensitivity pattern. The FWHM is $\sim 9'$ for
the 145 GHz PSBs; is $\sim 6'$ for the 245 GHz and 345 GHz
channels.}
\end{figure}
\begin{figure}
\plottwo {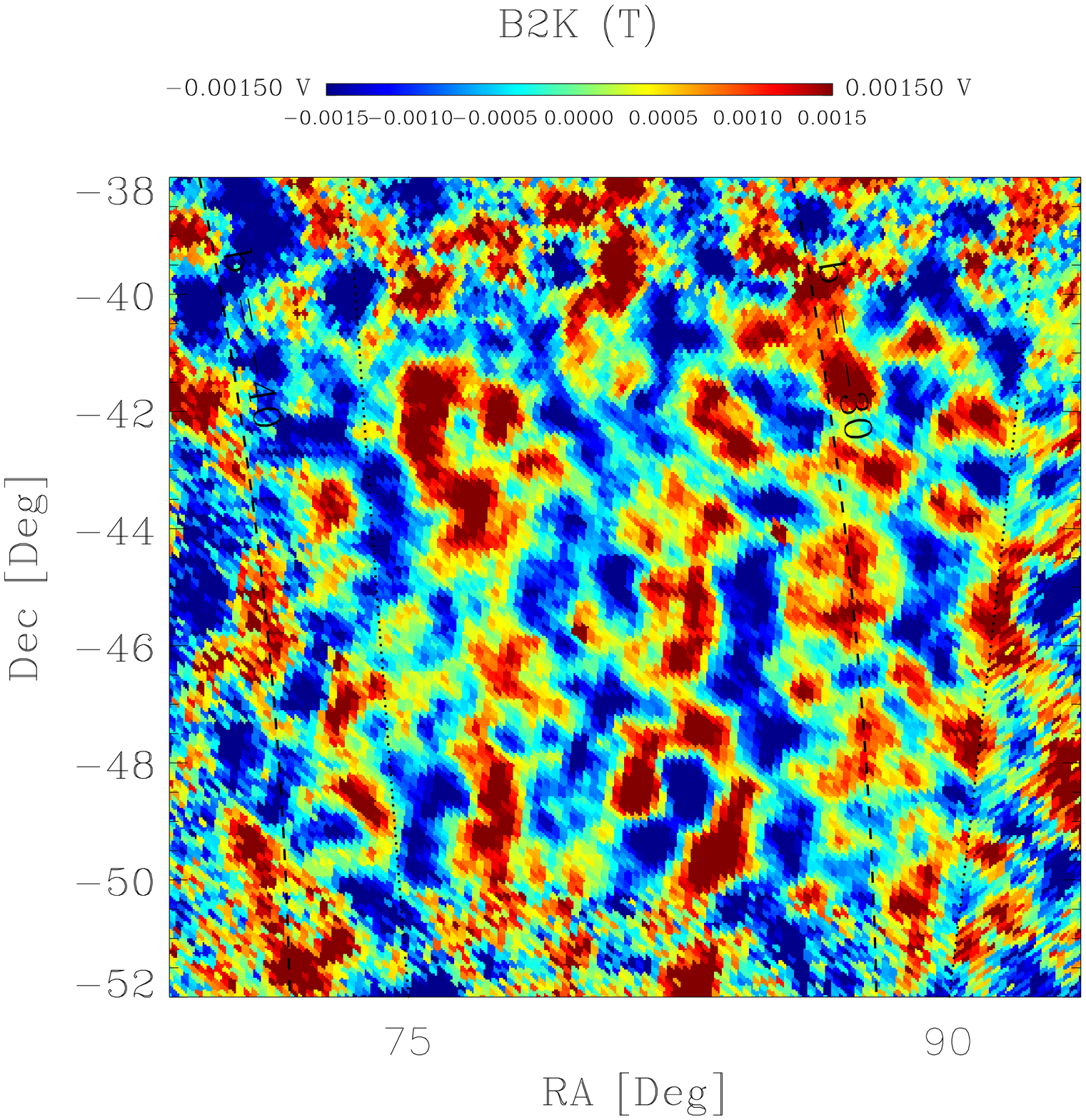} {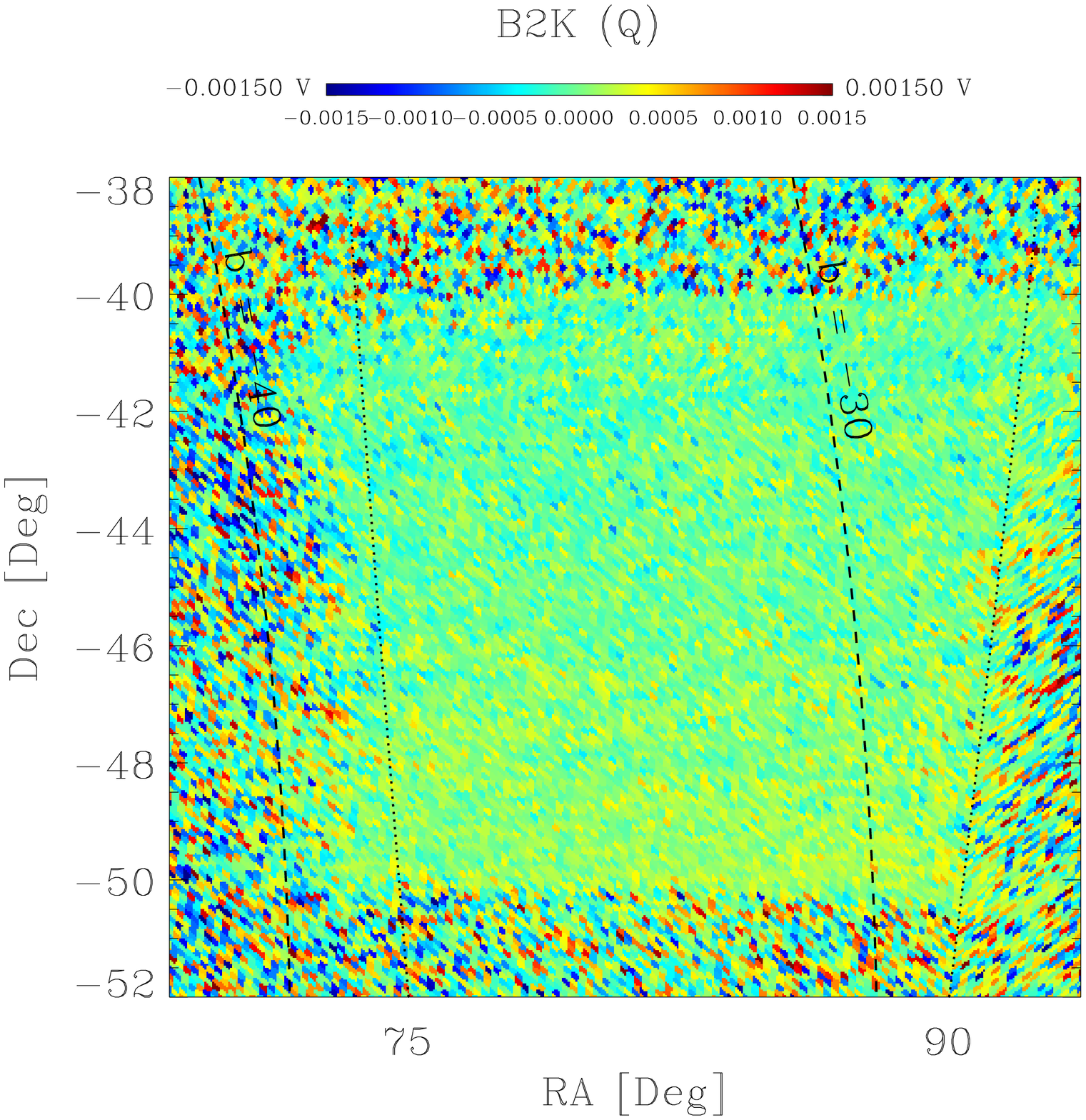} \caption{{\bf Left:}
Preliminary CMB anisotropy map in the region with deeper
integration surveyed during the B2K flight. The signals from the 8
PSB bolometers sensitive at 145 GHz have been averaged to obtain
the brightness map. The map has been obtained from the IGLS
optimal code (Natoli et al. 2001). The brightness units are not
absolutely calibrated and the pointing solution is preliminary.
{\bf Right:} Map of Stokes parameter Q obtained from the 8 PSB
bolometers, in the same units as the anisotropy map. Linear
polarization of the CMB can be extracted from the noise by means
of power spectrum analysis.}
\end{figure}
It has 8 pixels in two rows. The distance between pixels projected
in the sky is 0.5$^o$ both in azimuth and in elevation.  Each
pixel of the bottom row is a 145 GHz PSB with two independent
bolometers sensing the two orthogonal polarization directions
(Jones et al. 2003); each pixel on the top row senses a single
polarization direction in two frequencies (240 and 350 GHz). The
150 GHz beam is $9'$ FWHM, while the 240 and 350 GHz beams are
$\sim 6'$ FWHM. More details on the instrument can be found in
Montroy et al. (2003). The experiment has been designed to measure
the E modes (gradient) in the CMB polarization pattern. This has
been detected only by the DASI experiment, but the sensitivity was
not enough to constrain cosmological models more than with
anisotropy data. WMAP has published a detection of the TE cross
correlation from the first year of operation, consistent with the
adiabatic model inferred by the anisotropy measurement. B2K should
be able to measure the TE and EE power spectra at frequencies
higher than the DASI and WMAP ones (30 to 94 GHz), thus facing
different polarized foregrounds and nicely complementing them. The
B2K payload has been flown from the McMurdo base on Jan.7, 2003,
for a total of 11 days of operation in the stratosphere. In fig.4
we plot preliminary maps from B2K obtained from the PSBs at 145
GHz. We plan to re-fly B2K with an upgraded focal plane, to go
after the polarized foreground from cirrus dust and AGNs. This
information is essential for all the planned B-modes experiments
(e.g. BICEP, Dome-C etc.) and is very difficult to measure from
ground. The BOOMERanG optics can host an array of $\simgt 100$ PSB
at $\simgt 350$ GHz, providing a deep, high resolution survey of
polarized foreground emission at frequencies close to the ones
used for CMB polarization research.

 The BOOMERanG experiment is
supported in Italy by Agenzia Spaziale Italiana, Programma
Nazionale Ricerche in Antartide, Universita' di Roma La Sapienza;
by PPARC in the UK, by NASA, NSF OPP and NERSC in the U.S.A., and
by CIAR and NSERC in Canada.

\begin{quote}

\verb"Abroe, E., et al., 2003, submitted to Ap.J,"
astro-ph/0308355

\verb"Bennett, C., et al., 2003, Ap.J., in press,"
astro-ph/0302207

\verb"Crill, B., et al. 2003, Ap.J.S., 148, 527"

\verb"de Bernardis, P., et al., 2000, Nature, 404, 955"

\verb"de Bernardis, P., et al., 2002, Ap.J., 564, 559"

\verb"De Troia, G., et al., 2003, MNRAS, 343, 284"

\verb"Gorski, K. M., Hivon, E. and Wandelt, B.D., 1998, in"\\
\verb"Analysis Issues for Large CMB Data Sets, A. J. Banday," \\
\verb"R. K. Sheth and L. Da Costa eds., ESO, Ipskamp, NL, "\\
\verb"pgg. 37-42, see also"  http://www.eso.org/science/healpix/

\verb"Hivon, E., et al., 2003, in preparation"

\verb"Jones, W.C., et al., 2002," astro-ph/0209132

\verb"Lamarre, J.M., et al., 2003," astro-ph/0308075

\verb"Lange, A.E., et al., 2001, Phys.Rev., D63, 042001"

\verb"Masi, S., et al., 2001, et al., Ap.J, 553, L93"

\verb"Mauskopf, P., et al. 1997, Applied Optics, 36, 765"

\verb"Montroy, T., et al., 2003," astro-ph/0305593

\verb"Netterfield, B., et al., 2002, Ap.J., 571, 604"

\verb"Piacentini, F., et al. 2002, Ap.J.Suppl., 138, 315 "

\verb"Polenta, G., et al., 2002, Ap.J., 572, L27 "

\verb"Peebles, P.J.E, and Yu, J.T., 1970, Ap.J., 162, 815 "

\verb"Ruhl, J., et al., 2003, Ap.J. in press" ( astro-ph/0212229 )

\verb"Spergel, D.N., et al., 2003, Ap.J., in press,"
astro-ph/0302209

\verb"Sunyaev, R.A. & Zeldovich, Ya.B., 1970, Astr.SpaceSci., 7, 3
"

\end{quote}

\end{document}